\documentclass[amsmath,prb,twocolumn,showpacs,superscriptaddress]{revtex4}

\usepackage{amsfonts}
\usepackage{graphicx}

\begin{document}

\title{Nonequilibrium isolated molecule limit}
\date{\today}
\author{Michael Galperin}
\altaffiliation[Previous address: ]{Theoretical Division and Center for Integrated Nanotechnologies, Los Alamos National Laboratory, Los Alamos NM 87545, USA}
\affiliation{Department of Chemistry \& Biochemistry, University of California San Diego, La Jolla, CA 92093-0340, USA}
\author{Abraham Nitzan}
\affiliation{School of Chemistry, The Sackler Faculty of Sciences, Tel Aviv University, Tel Aviv 69978, Israel}
\author{Mark A. Ratner}
\affiliation{Department of Chemistry and Materials Research Center, Northwestern University, Evanston IL 60208, USA}

\begin{abstract}
Method developed by Sandalov and coworkers [Int. J. Quant. Chem. 
\textbf{94}, 113 (2003)] is applied to inelastic transport in the case
of strong correlations on the molecule, which is relatively weakly coupled 
to contacts. Ability of the approach to deal with the transport in the 
language of many-body molecular states as well as take into account
charge-specific normal modes and nonadiabatic couplings is stressed.
We demonstrate capabilities of the technique within simple model
calculations, and compare it to previously published approaches.  
\end{abstract}


\maketitle

\section{\label{intro}Introduction}
Development of experimental capabilities dealing with nanostructures brings
necessity of appropriate theoretical description of quantum transport 
(charge, spin, and heat) in mesoscopic junctions to the forefront of 
research.\cite{RatnerNitzan} 
Indeed, a lot of work has been done in this direction. 
In particular, many approaches are based
on the Landauer expression for current through such 
junctions in elastic tunneling regime.\cite{Landauer}
One of specific features of molecular transport junctions, 
the focus of molecular electronics, is flexibility of the molecules,
which results in inelastic features being much more pronounced in 
transport through such junctions as compared e.g. to semiconductor quantum dots.
Inelastic features are used as a diagnostic tool, helping to assure presence
of the molecule and study its characteristics in the junction within
inelastic electron tunneling spectroscopy both in the off-resonance
(IETS)\cite{StipeRezaeiHo_PRL} and resonant tunneling 
(RIETS)\cite{Zhitenev} situations.
Detailed discussion of the inelastic transport in molecular junctions
can be found in Refs.~\onlinecite{jpcm_review,science_review}.

Theoretical description of IETS is well established today both within simple 
models\cite{LakeDatta_prb45,LakeDatta_prb46,Ueba_ss493,Ueba_ss502503,Ueba_prb68,weak_elph} and more realistic 
calculations.\cite{LPLH_prl86,BLL_prl96,FBL_PRL93,PFB_nl6,LP_prl85,PCGSFG_nl4} 
Ability to predict quantitatively experimental findings is a sign
of maturity of the field. From theoretical perspective this success
is caused by ability to use well-established nonequilibrium perturbation 
(in electron-vibration coupling) technique. Indeed, in the off-resonant
situation electron-vibration coupling $M$ is an effectively small parameter,
$M\ll\sqrt{\Delta E^2 + (\Gamma/2)^2}$ with $\Delta E$ being resonant off-set
and $\Gamma$ characterizing strength of molecule-contacts coupling.
This allows expansion of the evolution operator in powers of $M$,
and truncation at low ($M^2$) order, the Born approximation, is usually
sufficient to get quantitatively correct predictions of IETS signal
in molecular junctions.

Resonant tunneling situation, $\Delta E=0$, provides richer physics.
While weak electron-vibration coupling, $M\ll\Gamma$, is treated within
perturbation theory also here,\cite{Hyldgaard,MitraAleinerMillis} 
the last fails in the opposite 
situation, $M>\Gamma$, when e.g. formation of polaron on the molecule
becomes possible. Theoretically this case till now mostly was treated either
within scattering theory (or isolated molecule) 
approach,\cite{WJW_prl,WJW_prb,BoncaTrugman,AlexandrovBratkovsky}
or within quasiclassical (rate or generalized rate equations)
scheme.\cite{MitraAleinerMillis,Datta_fseq,Datta_ph}
While the first treats electron-vibrational interaction (numerically)
exactly, it disregards Fermi populations in the contacts,
as well as dynamical features due to their presence,\cite{Flensberg}
and may lead to erroneous predictions.\cite{MitraAleinerMillis,strong_elph}
The last disregards quantum correlations, and as such is applicable
to either high temperature, $k_BT\gg\hbar\omega$, (truly classical)
situations, or (for generalized rate equations approach) quantum situations
when correlations in the system die much quicker than electron transfer 
(between contact and molecule) time ($\sim 1/\Gamma$).
Besides, these schemes lack formal procedure for improvement of their results
similar to taking into account higher order terms in perturbative expansion.
Recently we proposed nonequilibrium equation-of-motion (EOM) approach 
perturbative in molecule-contact coupling capable of dealing with
RIETS situation.\cite{strong_elph} 
The approach is formulated for simple resonant level 
model, but can be easily generalized for more realistic 
situations.\cite{HartleBeneshcThoss}
While it incorporates contacts (and hence nonequilibrium character of the
junction) into consideration, the price to pay is (generally) more
approximate level of description of electron-vibration interaction on 
the bridge. 
Alternatively, schemes exploring particular parameter regions
(slow vibration $\omega_0<\Gamma$,\cite{polaron,LD_prl,Kuznetsov}
small, $V\ll\omega_0$,\cite{MAM_prl} or big, $V\gg\omega_0$,\cite{MHM_prb}
bias) were proposed at a model level.

Another point especially important for resonant transport
(both elastic and inelastic), when actual oxidation or reduction of
the molecule takes place,
is necessity to speak in the language of many-body molecular states
contrary to single-particle molecular orbitals (the last is used in most
ab initio transport calculations today).
This includes electronic structure reorganization upon charging,
state dependent vibrational modes, anharmonicities, and non-Born-Oppenheimer
couplings. First schemes trying to treat transport in a many-body molecule
states language were recently proposed.\cite{Datta_fseq,Datta_ph,BS_condmat}

Difficulties in describing RIETS stem from absence of  well-established
{\em nonequilibrium atomic limit}\/ for molecule in the junction. 
Indeed, contacts play role of boundary conditions responsible
for establishing nonequilibrium state of the molecule. The last is
a complicated mixture of different charge (and excitation) states.
Approaches developed in molecular electronics community so far
either disregard boundary conditions and treat molecule as an
equilibrium object (scattering theory and isolated molecule treatments),
or establish this nonequilibrium state (mostly nonequilibrium Green function,
NEGF, approaches) being unable to map it into separate charge 
(or more exactly state) constituents. Note, density matrix based schemes
capable of such mapping were developed 
recently.\cite{Schoeller,RammerShelankovWabnig}
Such schemes however, miss time correlations, which may become important in
e.g. noise spectrum calculations. 

Hubbard operators is a natural language to talk about system (in our case 
subsystem - molecule) in terms of its states. Thus one seems to be
interested in utilizing nonequilibrium Hubbard operator Green function
technique for description of situations similar to RIETS, when ability
to establish nonequilibrium atomic limit of system is desirable. 
The approach should be capable to provide a systematic way
to take correlations into account (similar to perturbative
expansion in standard diagrammatic techniques). Such approach
originating from Kadanoff and Baym functional derivative EOM
scheme, was developed in the form of equilibrium Hubbard operator GFs
by Sandalov and coworkers for materials with strong electron
correlations (magnets with localized and partly localized moments,
Mott insulators, Kondo lattices, heavy fermion systems,
and high-Tc superconductors).\cite{Sandalov_ijqc} 
The method so far has been applied to model
elastic transport through quantum dots\cite{FES_prl,F_prb72,SN_prb75} 
and lowest states of double quantum 
dots,\cite{FES_pn,FE_jpcm,F_prb69,FE_prb70} and is completely ignored
in the molecular electronics community.

The goal of our present consideration is to
introduce inelastic transport description in the Coulomb blockade
regime within proper non-equilibrium atomic limit, and to attract
attention of the molecular electronics community to the proper
nonequilibrium approach capable of speaking in the language of 
many-particle states (rather than single-particle orbitals) and  
taking into account both molecular charge state dependent
normal modes (presently largely ignored in simulations) and non-adiabatic
couplings. 
Note, that Kondo physics is beyond the scope of current consideration.
The approach takes into account only on-the-molecule correlations
in a way that generalizes previous considerations.\cite{Datta_fseq,Datta_ph}
Note, that including many-body molecular states into consideration
of transport potentially allows for: 1.~Much more accurate molecular
structure simulation out of equilibrium than in current ab initio schemes,
due to possibility to employ equilibrium quantum chemistry methods
as a starting point for self-consistent procedure.
2.~Proper treatment of oxidation/reduction and corresponding electronic
and vibrational molecular structure changes, as well as non-adiabatic couplings,
3.~Ability to deal with general form of electron-vibration interaction
as long as it is localized in space (in the spirit of 
Ref.~\onlinecite{BoncaTrugman} but in
addition retaining many-body character of the junction),
4.~Calculation of noise spectrum of the junction due to preserved time
correlations (see e.g. Ref.~\onlinecite{noise} for detailed discussion),
5.~Proper treatment of degenerate situations due to preserved
space correlations (see e.g. Ref.~\onlinecite{Kenkre} for discussion).
Structure of the paper is the following. 
In Section~\ref{method} we briefly describe the method
in terms of many-electronic states of the system, and compare it to
previously proposed generalized master equation scheme.
Section~\ref{results} presents numerical examples of its application
to transport with discussion. Section~\ref{conclude} concludes.

\section{\label{method}Method}
Here we introduce model of molecular junction, briefly review
the basics of nonequilibrium Hubbard Green function technique, and
compare it to previously proposed generalized master equation approach.

\subsection{\label{model}Model}
As usual we consider molecular junction consisting of 3 parts: left (L) 
and right (R) contacts and the molecule (M). The contacts are assumed
to be reservoirs of free electrons each at its own equilibrium.
Molecule (or supermolecule if inclusion of parts of contacts is required) 
is the nonequilibrium part of the system. Besides, any external potential,
e.g. gate voltage probe, or additional contacts can be added to the picture 
if necessary. The Hamiltonian of the system is
\begin{equation}
 \label{H}
 \hat H = \hat H_L + \hat H_M + \hat H_R + \hat H_T
\end{equation}
where $\hat H_K$ ($K=L,R$) is Hamiltonian for contact $K$
\begin{equation}
 \label{HK}
 \hat H_K = \sum_{k\in K} \varepsilon_k \hat c_k^\dagger \hat c_k,
\end{equation}
$\hat H_M$ is Hamiltonian of the isolated molecule, and $\hat H_T$
is coupling between the subsystems
\begin{equation}
 \label{HT}
 \hat H_T = \sum_{k\in\{L,R\}; m\in M}\left(
  V_{km} \hat c_k^\dagger \hat d_m + V_{mk}\hat d_m^\dagger\hat c_k
 \right)
\end{equation}
$\hat d^\dagger$ ($\hat d$) and $\hat c^\dagger$ ($\hat c$) are 
creation (annihilation) operators for electron on the molecule and
in the contacts, respectively. Their indices $m$ and $k$ denote electronic
state in some chosen single-particle basis, and incorporate all the necessary
quantum indices (e.g. site and spin).

Now we want to consider molecular subsystem in the basis of
{\em many-electron states}\/ $|N,i>$, where $N$ stands for molecular
charge (number of electrons or excess electrons on the molecule)
and $i$ numerates different (e.g. excitation) states within the same
charge state block. Generally these states should not be orthonormal,
and consequences of overlap between different molecular states
(as well as overlap of molecular and contact states) were considered
in several papers.\cite{Sandalov_ijqc,FES_prb66} 
In what follows however we chose the states $|N,i>$ to be orthonormal
\begin{equation}
 <N,i|N',i'> = \delta_{N,N'}\delta_{i,i'}
\end{equation}  
in order to keep notation as simple as possible.
Hubbard operators are introduced as usual
\begin{equation}
 \label{X}
 \hat X_{(N,i;N',i')} \equiv |N,i><N',i'|
\end{equation}

In terms of these many-electron states transfer Hamiltonian becomes
\begin{equation}
 \label{HT_MB}
 \hat H_T = \sum_{k\in\{L,R\}; \mathcal{M}} \left(
 V_{k\mathcal{M}}\hat c_k^\dagger \hat X_\mathcal{M} +
 V_{\bar{\mathcal{M}}k} \hat X_{\mathcal{M}}^\dagger \hat c_k
 \right)
\end{equation}
where 
\begin{equation}
 \label{M}
 \mathcal{M}\equiv (N,i;N+1,j)
\end{equation}
denotes transition of the 
system from state $|N+1,j>$ to state $|N,i>$, while
$\bar{\mathcal{M}}\equiv (N+1,j;N,i)$ stands for the backward transition.
Transfer matrix element is 
\begin{equation}
 V_{k\mathcal{M}} \equiv \sum_{m\in M} V_{km}<N,i|\hat d_m|N+1,j>
\end{equation}
and $V_{\bar{\mathcal{M}}k}=V_{k\mathcal{M}}^{*}$.
Often many-electron states are chosen as eigenstates of isolated
molecule, in this case
\begin{equation}
 \label{HM_MB}
 \hat H_M = \sum_{|N,i>} E_{N,i} X_{(N,i;N,i)}
\end{equation}
with $E_{N,i}$ energies of the isolated molecule states.

\subsection{\label{current}Current expression}
Following derivation by Meir and Wingreen\cite{MeirWingreen,HaugJauho}
one gets usual expression for the current at interface $K=L,R$
\begin{align}
 \label{IK_t}
 I_K(t) =& \frac{e}{\hbar}\int_{-\infty}^t dt_1\, \mbox{Tr}\left[\right. 
 \nonumber \\
 & \Sigma_K^{<}(t,t_1)\, G^{>}(t_1,t) + G^{>}(t,t_1)\,\Sigma_K^{<}(t_1,t)
 \\ -&\left.
   \Sigma_K^{>}(t,t_1)\, G^{<}(t_1,t) - G^{<}(t,t_1)\, \Sigma_K^{>}(t_1,t)
 \right]
 \nonumber
\end{align}
which for steady-state situation simplifies to
\begin{equation}
 \label{IK_E}
 I_K = \frac{e}{\hbar} \int_{-\infty}^{+\infty}\frac{dE}{2\pi}\,
 \mbox{Tr}\left[\Sigma_K^{<}(E)\, G^{>}(E)-\Sigma_K^{>}(E)\, G^{<}(E)\right]
\end{equation} 
The only difference from the standard NEGF expression is that 
$\mbox{Tr}[\ldots]$ in (\ref{IK_t}) and (\ref{IK_E}) goes not
over single-electron basis, but over basis of {\em single-electron
transitions}\/ $\mathcal{M}$, Eq.(\ref{M}), 
between many-particle states of the molecule.

Self-energies $\Sigma_K$ in (\ref{IK_t}) and (\ref{IK_E}) are
defined on the Keldysh contour as
\begin{equation}
 \label{Sigma}
 \left[\Sigma_K (\tau,\tau')\right]_{\mathcal{M}\mathcal{M}'} \equiv
 \sum_{k\in K} V_{\bar{\mathcal{M}}k}\,g_k(\tau,\tau')\,
               V_{k\mathcal{M}'}
\end{equation}
with 
\begin{equation}
 \label{gk}
 g_k(\tau,\tau') \equiv -i < T_c\, \hat c_k(\tau)\,\hat c_k^\dagger(\tau') >
\end{equation}
GF for free electrons in the contacts. SEs projections are
\begin{align}
 \left[\Sigma_K^{<}(E)\right]_{\mathcal{M}\mathcal{M}'}
 &= i \Gamma^K_{\mathcal{M}\mathcal{M}'}(E) f_K(E)
 \\
 \left[\Sigma_K^{>}(E)\right]_{\mathcal{M}\mathcal{M}'}
 &= -i \Gamma^K_{\mathcal{M}\mathcal{M}'}(E) \left[1-f_K(E)\right]
\end{align}
where 
\begin{equation}
 \Gamma^K_{\mathcal{M}\mathcal{M}'}(E) \equiv \sum_{k\in K}
 V_{\bar{\mathcal{M}}k}\, V_{k\mathcal{M}'}\,\delta(E-\varepsilon_k)
\end{equation}
and $f_K(E)$ is the Fermi distribution in contact $K$.

GFs in (\ref{IK_t}) and (\ref{IK_E}) are Hubbard operator GFs defined on 
the Keldysh contour as
\begin{equation}
 \label{GMM}
 G_{\mathcal{M}\mathcal{M'}}(\tau,\tau') \equiv
 -i <T_c\, \hat X_\mathcal{M}(\tau)\,
           \hat X_{\mathcal{M}'}^\dagger(\tau')>
\end{equation}
Note, that operators in (\ref{gk}) and (\ref{GMM}) are in Heisenberg 
representation. Note also, that $\mathcal{M}$ and $\mathcal{M}'$ in (\ref{GMM})
may be (in principle) arbitrarily far away from one another in 
the charge space. GF (\ref{GMM}) represents correlation between 
different single-electron molecular many-body state transitions
due to coupling to the same bath (contacts). In practice however,
it seems unreasonable to go beyond correlations between nearest
charge space blocks.

In order to show connection of the present formalism to previously
proposed generalized rate equation (master equation in the Fock space)
approach, we have to realize that the last misses correlations
both in space and time. So to reduce present GF description to the
master equation in the Fock space, we need to make several simplifications:
\begin{enumerate}
\item {\em Diagonal approximation.} We have to stick to
 diagonal elements of GF only $G_{\mathcal{M}\mathcal{M}}$
 with $\mathcal{M}=(N,i;N+1,j)$.
\item {\em Markov approximation.} We have to consider only GFs
 of equal times $G(t,t)$. In order to reduce GFs of different times 
 entering (\ref{IK_E}) to equal times quantities 
 we use approximation
 \begin{equation}
  G_{\mathcal{M}\mathcal{M}}(t-t') \approx 
  \exp[i\Delta^{0}_\mathcal{M}(t'-t)]\,
  G_{\mathcal{M}\mathcal{M}}(t-t)
 \end{equation}
 where 
 \begin{equation}
  \label{Delta0}
  \Delta^{0}_\mathcal{M} \equiv E_{N+1,j}-E_{N,i}
 \end{equation}
\end{enumerate} 
Now, noting that
\begin{align}
 iG^{>}_{\mathcal{M}\mathcal{M}}(t-t) &= P^N_i
 \\
-iG^{<}_{\mathcal{M}\mathcal{M}}(t-t) &= P^{N+1}_j
\end{align}
are probabilities to find molecule in state $|N,i>$ and $|N+1,j>$
respectively, we get from (\ref{IK_E})
\begin{align}
 I_K = \frac{e}{\hbar}\sum_{N;i,j}&\left(
  \Gamma^K_{(N,i;N+1,j)}\, f_K(E_j^{N+1}-E_i^N)\, P^N_i
 \right. \\ &  \left.
 -\Gamma^K_{(N,i;N+1,j)}\,[1-f_K(E^{N+1}_j-E^N_i)]\, P^{N+1}_j
 \right)
 \nonumber
\end{align}
If now one restricts attention only to particular charge space
block $N_0$ and its nearest neighbors, one gets Eqs.~(6) and (7)
of Ref.~\onlinecite{Datta_fseq}.

\subsection{\label{EOM}General equation for GF}
Now, when expression for the current is established, we need a procedure
to calculate Hubbard operators GF (\ref{GMM}). Note, that standard 
diagrammatic techniques are inapplicable here, due to lack of the 
Wick's theorem (since Hubbard operators are many-particle operators).
An alternative to diagrammatic expansion in the form of functional 
derivative equation-of-motion technique was developed in 
Ref.~\onlinecite{Sandalov_ijqc}.
Here we briefly review steps needed to obtain EOM for GF we will use
in our numerical simulations. 

Following Ref.~\onlinecite{Sandalov_ijqc} we start by writing EOM
for Hubbard operator $\hat X_\mathcal{M}(\tau)$, 
where $\mathcal{M}\equiv (N,i;N+1,j)$. This leads to
\begin{align}
 \label{X_EOM}
 &\left[i\frac{\partial}{\partial\tau}-\Delta_\mathcal{M}^{0}\right]
 \hat X_\mathcal{M}(\tau) = 
 \nonumber \\
 &\sum_{k\in\{L,R\};\ell}\left(
 -V_{k(N+1,j;N+2,\ell)}\,\hat c_k^\dagger(\tau)\,\hat X_{(N,i;N+2,\ell)}(\tau)
 \right.\nonumber\\
 &-V_{k(N-1,\ell;N,i)}\,\hat c_k^\dagger(\tau)\,\hat X_{(N-1,\ell;N+1,j)}(\tau)
 \\
 &+V_{(N+1,j;N,\ell)k}\,\hat X_{(N,i;N,\ell)}(\tau)\,\hat c_k(\tau)
 \nonumber \\ &\left.
 +V_{(N+1,\ell;N,i)k}\,\hat X_{(N+1,\ell;N+1,j)}(\tau)\,\hat c_k(\tau)
 \right)
 \nonumber
\end{align}
In what follows we disregard first $2$ terms on the right-hand-side,
since they describe simultaneous transfer of $2$ electrons between
contact and molecule, which is beyond the scope of this consideration.
It is clear that when writing EOM for GF (\ref{GMM}) terms in 
the right-hand-side of (\ref{X_EOM}) will produce correlation
functions of the form
\begin{equation}
 \label{corr}
 <T_c\,\hat X_\xi(\tau)\,\hat c_k(\tau)\,\hat X_{\mathcal{M}'}^\dagger(\tau')>
\end{equation}
which can not be factorized into product of single-excitation GF (\ref{GMM})
and contact single-electron GF (\ref{gk}) due to lack of the Wick's theorem.

In order to make this separation a trick with auxiliary fields 
$\mathcal{U}_\xi(\tau)$
is employed. We need to introduce additional disturbance potential
\begin{equation}
 \label{disturb}
 \hat H_\mathcal{U}(\tau) \equiv \sum_{N;i,j}
  \mathcal{U}_{(N,i;N,j)}(\tau)\,\hat X_{(N,i;N,j)}(\tau)
\end{equation}
and corresponding generating functional
\begin{equation}
 \label{generat}
 \hat S_\mathcal{U} \equiv 
 \exp\left[-i\int_c d\tau\,\hat H_\mathcal{U}(\tau)\right]
\end{equation}
Then defining GF of $2$ arbitrary operators $\hat A$ and $\hat B$ 
in the presence of auxiliary fields $\mathcal{U}$
\begin{align}
 \label{GAB}
 G_{AB}(\tau,\tau') &\equiv 
 -i<T_c\,\hat A(\tau)\,\hat B(\tau')>_\mathcal{U}
 \\ &\equiv
 -i\frac{<T_c\,\hat S_\mathcal{U}\,\hat A(\tau)\,\hat B(\tau')>}
        {<T_c\,\hat S_\mathcal{U}>}
 \nonumber
\end{align}
one easily can get the following identity
\begin{align}
 \label{trick}
 &-i<T_c\,\hat X_\xi(\tau'')\,\hat A(\tau)\,\hat B(\tau')>_\mathcal{U}
\\
 &= \left[<T_c\,\hat X_\xi(\tau'')>_\mathcal{U}+
        i\frac{\delta}{\delta\mathcal{U}_\xi(\tau'')}\right] G_{AB}(\tau,\tau')
 \nonumber
\end{align}
Eq.(\ref{trick}) allows to express correlation function (\ref{corr})
in terms of single-excitation GF and its functional derivatives
relative to auxiliary fields. Note, that putting (at the end) auxiliary
fields to be zero turns (\ref{GAB}) into a standard definition of GF.

So, introducing auxiliary fields as in Eqs.~(\ref{disturb}) and (\ref{generat})
and using expression (\ref{trick}), one gets general EOM for Hubbard operator
GF (\ref{GMM}) in the form
\begin{align}
 \label{GMM_fderiv}
 &\left[i\frac{\partial}{\partial\tau}-\Delta_\mathcal{M}^{0}\right]
  G_{\mathcal{M}\mathcal{M}'}(\tau,\tau') 
 \nonumber \\
 &-\sum_\ell\left(\mathcal{U}_{(N+1,j;N+1,\ell)}(\tau)\,
                  G_{(N,i;N+1,\ell)\mathcal{M}'}(\tau,\tau')
 \right. \nonumber \\ &\qquad \left.
                 -\mathcal{U}_{(N,\ell;N,i)}(\tau)\,
                  G_{(N,\ell;N+1,j)}(\tau,\tau')\right)
 \nonumber \\
 &= \delta(\tau,\tau')\,P_{\mathcal{M}\mathcal{M}'}(\tau)
 \nonumber \\
 &+\sum_\ell\left(
 \left[<T_c\,\hat X_{(N,i;N,\ell)}(\tau)>_\mathcal{U}
      +i\frac{\delta}{\delta\mathcal{U}_{(N,i;N,\ell)}(\tau)}\right]
 \right.  \\ &
 \qquad\times \sum_{\mathcal{M}''}\int_c d\tau''\,
      \Sigma_{(N,\ell;N+1,j)\mathcal{M}''}(\tau,\tau'')\,
      G_{\mathcal{M}''\mathcal{M}'}(\tau'',\tau')
 \nonumber \\ & +
 \left[<T_c\,\hat X_{(N+1,\ell;N+1,j)}(\tau)>_\mathcal{U}
      +i\frac{\delta}{\delta\mathcal{U}_{(N+1,\ell;N+1,j)}(\tau)}\right]
 \nonumber \\ & \left.
 \qquad\times \sum_{\mathcal{M}''}\int_c d\tau''\,
      \Sigma_{(N,i;N+1,\ell)\mathcal{M}''}(\tau,\tau'')\,
      G_{\mathcal{M}''\mathcal{M}'}(\tau'',\tau')
 \right)
 \nonumber
\end{align}
where $\Delta_\mathcal{M}^{0}$ is defined in (\ref{Delta0}) and
\begin{equation}
 \label{PMM}
 P_{\mathcal{M}\mathcal{M}'} \equiv
 <T_c\,\hat X_{(N,i;N,i')}(\tau)+\hat X_{(N+1,j';N+1,j)}(\tau)>_\mathcal{U}
\end{equation}
Eq.(\ref{GMM_fderiv}) is a general equation for Hubbard operator GF,
representing an alternative to standard diagrammatic technique approaches.
Role of expansion in small parameter here play functional derivatives
in auxiliary fields. Level of approximation is defined by order
of the derivative used in evaluation of GF. At the end of differentiations
auxiliary fields are put to be zero, and resulting expression is 
equation for GF at particular level of approximation.

\subsection{\label{loop}First loop approximation}
The simplest approximation, Hubbard I (HI), is obtained from
(\ref{GMM_fderiv}) by keeping only diagonal averages, omitting all functional
derivatives, and $\mathcal{U}\to 0$
\begin{align}
 \label{HI}
 &\left[i\frac{\partial}{\partial\tau}-\Delta_\mathcal{M}^{0}\right]
  G_{\mathcal{M}\mathcal{M}'}(\tau,\tau')
  = \delta(\tau,\tau')\,\delta_{\mathcal{M}\mathcal{M}'}\,P_\mathcal{M}
 \\
 &+P_\mathcal{M}\sum_{\mathcal{M}''}\int_c d\tau'' \,
   \Sigma_{\mathcal{M}\mathcal{M}''}(\tau,\tau'')\,
   G_{\mathcal{M}''\mathcal{M}'}(\tau'',\tau')
 \nonumber
\end{align}

Following most of the papers, employed the method so 
far,\cite{FES_prl,F_prb72,SN_prb75}
in our consideration we go one step further - we take one functional 
derivative to get so called first loop approximation.
Note, here we take derivative of the GF only, disregarding fluctuations
of the spectral weight $P$. After performing the differentiation
once more we keep only diagonal averages, omit all functional derivatives,
and $\mathcal{U}\to 0$. Since the procedure was described in details in 
many papers (see e.g. Refs.~\onlinecite{Sandalov_ijqc,SN_prb75}), 
here we present only final result
\begin{align}
 \label{firstloop}
 &\left[i\frac{\partial}{\partial\tau}-\Delta_\mathcal{M}^{0}\right]
  G_{\mathcal{M}\mathcal{M}'}(\tau,\tau')
 -i\sum_{\kappa,\ell}\sum_{\mathcal{M}''}\int_c d\tau''
 \nonumber \\
 &\left[ \Sigma_{(N,\kappa;N+1,j)\mathcal{M}''}(\tau,\tau'')\,
         D_{\mathcal{M}''(N-1,\ell;N,i)}(\tau'',\tau+)\,
 \right. \nonumber \\ &\qquad \times
         G_{(N-1,\ell;N,\kappa)\mathcal{M}'}(\tau,\tau')
 \nonumber \\
 &      -\Sigma_{(N,\kappa;N+1,j)\mathcal{M}''}(\tau,\tau'')\,
         D_{\mathcal{M}''(N,\kappa;N+1,\ell)}(\tau'',\tau+)\,
 \nonumber \\ &\qquad \times
         G_{(N,i;N+1,\ell)\mathcal{M}'}(\tau,\tau')
 \nonumber \\
 &      +\Sigma_{(N,i;N+1,\kappa)\mathcal{M}''}(\tau,\tau'')\,
         D_{\mathcal{M}''(N,\ell;N+1,\kappa)}(\tau'',\tau+)\,
 \nonumber \\ &\qquad \times
         G_{(N,\ell;N+1,j)\mathcal{M}'}(\tau,\tau')
 \\
 &      -\Sigma_{(N,i;N+1,\kappa)\mathcal{M}''}(\tau,\tau'')\,
         D_{\mathcal{M}''(N+1,j;N+2,\ell)}(\tau'',\tau+)\,
 \nonumber \\ &\qquad \times \left.
         G_{(N+1,\kappa;N+2,\ell)\mathcal{M}'}(\tau,\tau') \right]
 \nonumber \\
 & = \delta(\tau,\tau')\,\delta_{\mathcal{M}\mathcal{M}'}\,P_\mathcal{M}
 \nonumber \\
 &+P_\mathcal{M}\sum_{\mathcal{M}''}\int_c d\tau'' \,
   \Sigma_{\mathcal{M}\mathcal{M}''}(\tau,\tau'')\,
   G_{\mathcal{M}''\mathcal{M}'}(\tau'',\tau')
 \nonumber
\end{align}
where $\mathcal{M}\equiv (N,i;N+1,j)$ and $D$ is so called full locator,
which (in the first loop approximation) obeys the same equation 
(\ref{firstloop}) as GF but without spectral weight $P_\mathcal{M}$
multiplying delta function in the right-hand-side.

Expressions for GFs $G$ and $D$  (first loop approximation) in the shorthand
(matrix in both Fock space and Keldysh contour variables) 
notation can be written as
\begin{align}
 \label{shorthand_G}
 \hat D^{-1}\, G &= P \\
 \label{shorthand_D}
 \hat D^{-1}\, D &= 1
\end{align}
where
\begin{align}
 \label{D_1}
 \hat D^{-1}  &\equiv 
 \left[i\frac{\partial}{\partial\tau}-\Delta_\mathcal{M}
 - P\,\Sigma \right]
 \\
 \Delta_\mathcal{M} &=\Delta_\mathcal{M}^{0}+\delta\Delta_\mathcal{M}
\end{align}
and $\delta\Delta_\mathcal{M}$ is given by the second term in
the left-hand-side of Eq.(\ref{firstloop}), it is responsible for
shifts of transition energies in the molecule due to contacts
induced correlation. One sees that Eq.(\ref{shorthand_D}) has
usual structure of the Dyson equation, which is obtained in 
standard diagrammatic expansion. The only difference is dressing
of SE $\Sigma$ by spectral weight $P$. Thus formally one can use all
the standard equations, using dressed SE $\Sigma$ everywhere, 
to get desired projections of GF $D$. When $D$ is known, $G$ is obtained
by simple matrix multiplication
\begin{equation}
 \label{GDP}
 G = D\, P
\end{equation}  
Note, side of matrix dressing by spectral weight $P$ is different for 
$\Sigma$ and $G$, compare Eqs.~(\ref{D_1}) and (\ref{GDP}).
Note also, that the scheme is self-consistent, since both
transition energies shift $\delta\Delta$ and spectral weights $P$
depend on GF $G$, while the last depends on these quantities.
In particular ($\mathcal{M}\equiv (N,i;N+1,j)$)
\begin{align}
 &P_\mathcal{M} = N_{N,i} + N_{N+1,j} &
 \\
 &N_{N,i} \equiv <\hat X_{(N,i;N,i)}> 
 = iG^{>}_{\mathcal{M}\mathcal{M}}(t,t)
 &\mbox{for any j} 
 \\
 &N_{N+1,j} \equiv <\hat X_{(N+1,j;N+1,j)}> 
 = -iG^{<}_{\mathcal{M}\mathcal{M}}(t,t)
 &\mbox{for any i}
\end{align}
Here $N_{N,i}$ and $N_{N+1,j}$ are probabilities to find system
in the state $|N,i>$ and $|N+1,j>$ respectively.

\section{\label{results}Numerical results and discussion}
Here we present results of simulations within first loop approximation.
In order to speed up calculations we employed also diagonal 
approximation.\footnote{Note, that approximation like the one presented 
in Section~\ref{loop} may result
in unphysical behavior. In particular, retarded and advanced SEs and GFs are
not Hermitian conjugates of one another. While the issue does not arise
in the diagonal approximation implemented for calculations here,
this is a problem for a more general consideration. A simple workaround
is to use average of two Dyson-like expressions: one with $\hat D^{-1}$,
Eq.(\ref{D_1}), applied from the left and one from the right.
This leads to set of equations for GFs, which under Markov approximation
reduce to widely employed equations for DM in dissipative environment.} 
We consider transport through quantum dot and double quantum dot,
discuss obtained data, and compare it to previously published results.

\subsection{\label{QD}Quantum dot}
In the case of quantum dot molecular Hamiltonian is
\begin{equation}
 \hat H_M = \sum_{\sigma=\{\uparrow,\downarrow\}}
 \varepsilon_\sigma\hat n_\sigma
 + U \hat n_\uparrow \hat n_\downarrow
\end{equation}
where $\sigma$ indicates spin projection and 
$\hat n_\sigma=\hat d_\sigma^\dagger\hat d_\sigma$.
Full Fock space of the molecular part of the
system (without vibrations) consists of one empty state ($|0>\equiv |0,0>$), 
two single-electron states,
($|\uparrow>\equiv |1,\uparrow>$ and $|\downarrow>\equiv |1,\downarrow>$), 
and one doubly occupied state ($|2>\equiv |2,0>$). 
Transitions between these states to be considered are spin up electron
transfers ($0\uparrow$ and $\downarrow 2$) and spin down
electron transfers ($0\downarrow$ and $\uparrow 2$). 
Writing Eq.(\ref{firstloop}) in the basis of these transitions
one gets equations obtained in Ref.~\onlinecite{F_prb72}.

\begin{figure}[htbp]
\centering\includegraphics[width=\linewidth]{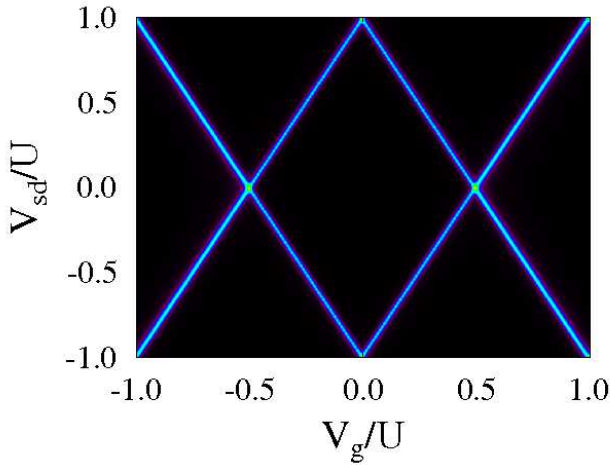}
\caption{\label{fig_qd}
(Color online) Conductance map for elastic transport through quantum dot. 
See text for parameters.
}
\end{figure}

Figure~\ref{fig_qd} presents conductance map for elastic transport through 
quantum dot. Parameters of the calculation are $T=10$~K, 
$\varepsilon_\sigma=-0.5$~eV, $\Gamma^K_\sigma=0.01$~eV 
($\sigma=\uparrow,\downarrow$ and $K=L,R$), and $U=1$~eV.
As usually one has areas of blockaded transport (inside part of diamonds)
with fixed population on the dot ($0$, $1$, and $2$ from right to left),
and transition areas (between the diamonds) where population on the dot
is noninteger (see e.g. Ref.~\onlinecite{CBKondo} for more detailed discussion).

In order to treat inelastic transport we add to $\hat H_M$ molecular vibration
linearly coupled to electron(s) on the dot
\begin{equation}
\label{HM_qd_vib_x}
 \omega_0\hat a^\dagger\hat a + 
 M(\hat a+\hat a^\dagger)\sum_\sigma \hat n_\sigma
\end{equation}
Such model is frequently used to describe inelastic transport in
molecular junctions. In a sense it is similar to Marcus theory,
and describes shift of the molecular vibration when molecule is charged 
due to electron transfer from/to the contacts. In general,
model with non-diagonal electron-vibration coupling also can be
considered within the formalism. 

After small polaron (Lang-Firsov or canonical) transformation\cite{Mahan}
linear coupling 
term is eliminated, while energy level position $\varepsilon_\sigma$ and 
Hubbard repulsion $U$ are renormalized 
($\varepsilon_\sigma\to\varepsilon_\sigma-M^2/\omega_0$ and 
$U\to U-2M^2/\omega_0$), and transfer matrix elements in $\hat H_T$,
Eq.(\ref{HT}), are dressed with shift operators 
($\hat d_\sigma\to \hat d_\sigma\hat{\mathcal{X}}$)
\begin{equation}
 \label{Xvib}
 \hat{\mathcal{X}} = \exp\left[-\lambda (\hat a^\dagger-\hat a)\right]
\end{equation}
where $\lambda=M/\omega_0$. In what follows we disregard renormalization
of $\varepsilon_\sigma$ and $U$, assuming that it was included in 
definition of these parameters. 

Now molecule is characterized by direct product
of electronic and vibrational spaces, so its state should be 
indicated by additional index $v$ showing state of the vibration,
i.e. molecular subspace is spanned by states 
$|0,v>$, $|\uparrow,v>$, $|\downarrow,v>$, and $|2,v>$, 
where $v\in\{0,1,2,3,\ldots\}$. One has to consider the same electronic
transitions as in the case of elastic transport, but in addition
all possible transitions between states of the vibration have to be 
included. Transitions between these states (within the model) are
possible only by electron transfer between molecule and contacts.
Due to shift operators (\ref{Xvib}) appearing in $\hat H_T$
SEs (\ref{Sigma}) are now dressed with corresponding vibrational
overlap integrals ($\mathcal{M}\equiv(N,i,v_i;N+1,j,v_j)$)
\begin{equation}
 \label{ovlp_x}
 \Sigma_{\mathcal{M}\mathcal{M'}} \to
 \Sigma_{\mathcal{M}\mathcal{M'}} \times
 <v_i|\hat{\mathcal{X}}|v_j>\, <v_i'|\hat{\mathcal{X}}|v_j'>
\end{equation} 
with
\begin{align}
 &<v|\hat{\mathcal{X}}|v'> = e^{-\lambda^2/2} (-1)^{(v-v')\theta(v-v')}
 \lambda^{v_{max}-v_{min}} 
 \\ &\qquad \times
 \left[\frac{v_{min}!}{v_{max}!}\right]^{1/2}
 L_{v_{min}}^{v_{max}-v_{min}}(\lambda^2)  
 \nonumber
\end{align}
where $v_{min}$ ($v_{max}$) is minimal (maximal) of $v$ and $v'$,
$\theta(x)$ is step function, and $L_n^m$ is Laguerre polynomial.
Note an important formal difference between the present approach
and the one presented in Ref.~\onlinecite{CBKondo}. While in the last 
we had to consider separately electron and phonon dynamics, which
leads to convolution of electron GF (electron dynamics) with 
Franck-Condon factors (phonon dynamics), here the situation is
different. Since we consider generalized Fock space (product of 
electronic and vibrational ones), within the formalism 
strictly speaking we do not have inelastic processes at all.
Instead one has to consider elastic scattering events between
electron-vibrational states. As a result the role played previously by
the Franck-Condon factors (to introduce vibrational dynamics)
now is included into Hubbard GF of the generalized Fock space.  

\begin{figure}[htbp]
\centering\includegraphics[width=\linewidth]{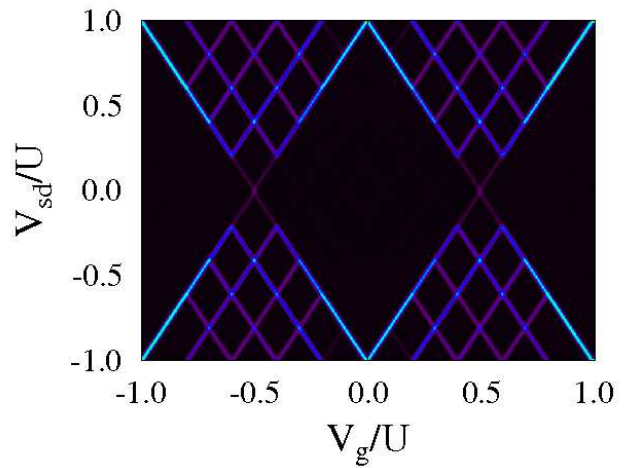}
\caption{\label{fig_vib_x}
(Color online) Conductance map for inelastic transport through quantum dot.
Linear coupling model (\ref{HM_qd_vib_x}). See text for parameters.
}
\end{figure}

Figure~\ref{fig_vib_x} presents conductance map for inelastic transport
through quantum dot within linear coupling model (\ref{HM_qd_vib_x}). 
Parameters of the calculation are $\omega_0=0.2$~eV and $M=0.4$~eV,
all the other parameters are as in Fig.~\ref{fig_qd}. 
Within the calculation we restricted  vibrational subspace to 4 lowest 
levels ($v\in\{0,1,2,3\}$). As expected, besides elastic peaks in the
conductance map we get additional resonant vibrational features
corresponding to inelastic processes. This Figure is equivalent
to Fig.~3a of Ref.~\onlinecite{CBKondo}, where calculation was done within
perturbative (in coupling to electrodes) nonequilibrium EOM
approach for the same model. As previously,\cite{CBKondo} 
within the model (see also discussion below) distance between the 
diamond edges (elastic peak) and vibrational sidebands is defined
by the oscillator frequency. Increase in electron-vibrational coupling
would result in both more pronounced vibrational features and suppression
of transport in the low source-drain voltage region due to Franck-Condon 
blockade. While increase in temperature would produce also vibrational 
sidebands corresponding to phonon absorption (features inside the diamond). 

Finally, we want to demonstrate capabilities of the present scheme, which go 
beyond approaches previously used to treat inelastic transport. 
Suppose our molecule is small enough, so that upon charging it changes 
its normal modes essentially. Suppose also, that from all the normal modes
of the molecule only one is coupled to tunneling electron.
Inelastic transport in this case can be modeled by assigning
different vibration frequencies to different charge states of the
molecule. In our quantum dot model this corresponds to situation,
when vibrational frequencies for $|0,v>$, $|\sigma,v>$, and $|2,v>$
states are different - $\omega_0^{(0)}$, $\omega_0^{(1)}$, 
and $\omega_0^{(2)}$ respectively. Self-energies due to electron transfer
between molecule and contacts once more have to be dressed by  
overlap integrals between different vibrational wavefunctions,
as is shown in Eq.(\ref{ovlp_x}).
However this time (when vibrational frequencies change) 
the integrals should be calculated in the way discussed in 
Refs.~\onlinecite{Ruhoff,RuhoffRatner}  

\begin{figure}[htbp]
\centering\includegraphics[width=\linewidth]{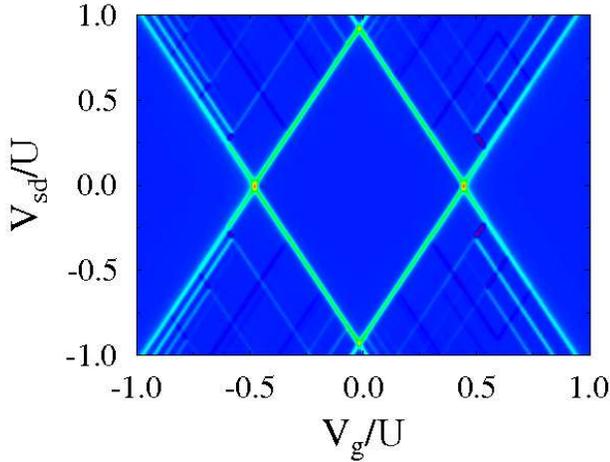}
\caption{\label{fig_vib}
(Color online) Conductance map for inelastic transport through quantum dot.
Charge state dependent frequencies. See text for parameters.
}
\end{figure}

Figure~\ref{fig_vib} presents conductance map for inelastic transport 
through quantum dot, when vibrational frequency depends on charge state
of the dot. Parameters of the calculation are $\omega_0^{(0)}=0.2$~eV,
$\omega_0^{(1)}=0.3$~eV, $\omega_0^{(2)}=0.25$~eV and shift vector
for both transitions was taken to be $0.5$~\AA 
(see Refs.~\onlinecite{Ruhoff,RuhoffRatner} for
detailed explanation), all the other parameters are as in Fig.~\ref{fig_qd}
One sees that result of calculation is counter intuitive at the first sight.
Naively one could expect to see inelastic peaks at each diamond edge
(each charge state of the quantum dot)
being separated by frequency corresponding to the neighboring 
charge state (RIETS probes frequencies of the intermediate ion).
Real picture is more complicated however. Let consider electron transfer
between $2$ particular charge states of the quantum dot, say between 
states $|0,v_0>$ and $|\sigma,v_1>$ upon electron transfer from contact
to molecule. In this case change in the subsystem energy, which will be 
observed in transport as inelastic peak in conductance, is
$v_1\omega_0^{(1)}-v_0\omega_0^{(0)}$ (we omit here change in elastic 
electronic energy for simplicity, this will define only position of the
elastic peak in conductance). Since $v_0$ and $v_1$ in principle can be any
non-negative numbers, it is clear that one can observe a a progression
of frequencies. Note, that in this progression one can see inelastic peaks
in conductance, separated from the elastic one by frequency which does not
exist in the system at all (e.g. $\omega_0^{(1)}-\omega_0^{(0)}$). 
Note also, that due to overlap factors involved the lowest frequencies
of the progression will be observed better in RIETS signal.     
Non-adiabatic couplings can be included in calculation in a similar way.

\subsection{\label{DQD}Double quantum dot}
Molecular Hamiltonian for double quantum dot is
\begin{align}
 \hat H_M &= \sum_{i=\{1,2\}\,\sigma=\{\uparrow,\downarrow\}}
 \varepsilon_{i\sigma} \hat n_{i\sigma}
 - t_{12,\sigma}\left(\hat d_{1\sigma}^\dagger\hat d_{2\sigma}
                     +\hat d_{2\sigma}^\dagger\hat d_{1\sigma}\right) 
 \\
 &+ \sum_i U_i\hat n_{i\uparrow}\hat n_{i\downarrow}
  + U_{12} \hat n_1 \hat n_2
 \nonumber
\end{align}
where $i=\{1,2\}$ numbers sites and $\sigma=\{\uparrow,\downarrow\}$
stands for spin projection, $\hat d^\dagger$ ($\hat d$) is
creation (annihilation) operator, 
$\hat n_{i\sigma}=\hat d_{i\sigma}^\dagger\hat d_{i\sigma}$,
and $\hat n_i=\hat n_{i\uparrow}+\hat n_{i\downarrow}$.
We assume that site $1$ is coupled to the left contact, while 
site $2$ - to the right. 

We chose many-body states for molecular subsystem in the form
$|1\uparrow,1\downarrow,2\uparrow,2\downarrow>$.
Unlike choice of Refs.~\onlinecite{FES_pn,FE_jpcm,F_prb69,FE_prb70} 
these are not eigenstates of the
molecular Hamiltonian. As a result EOM for Hubbard operator GFs couples
them also by hopping $t_{12,\sigma}$. Besides this all the treatment
presented in Section~\ref{method} remains the same. There are $16$ states
($1$, $4$, $6$, $4$, and $1$ states for $0$, $1$, $2$, $3$, and $4$ electrons
in the system respectively) and $32$ single-electron transitions 
($16$ for each spin block) to be considered. 

\begin{figure}[htbp]
\centering\includegraphics[width=\linewidth]{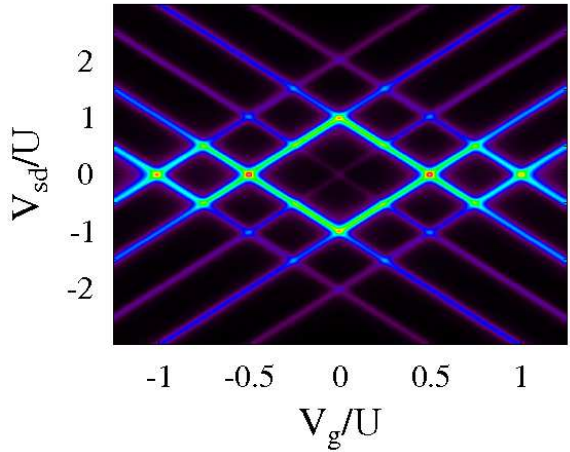}
\caption{\label{fig_dqd}
(Color online) Conductance map for elastic transport through double quantum
dot. See text for parameters.
}
\end{figure}

\begin{figure}[htbp]
\centering\includegraphics[width=0.8\linewidth]{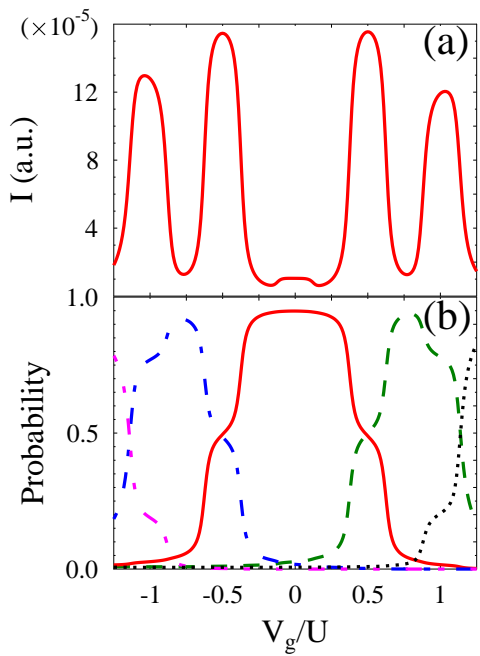}
\caption{\label{fig_dqd_cut}
(Color online) Elastic transport quantum dot at fixed source-drain
voltage $V_{sd}/U=0.25$:
(a) Current vs. gate voltage, (b) Probability to find double quantum dot empty 
(dotted line, black), singly- (dashed line, green), doubly- (solid line, red),
triply- (dash-dotted line, blue), or fully-occupied (dash-double-dotted line,
magenta). Parameters are the same as in Fig.~\ref{fig_dqd}.
}
\end{figure}

Figure~\ref{fig_dqd} shows conductance map for elastic transport through
double quantum dot. Parameters of the calculation are $T=10$~K,
$\varepsilon_{i\sigma}=-0.5$~eV, $t_{12,\sigma}=0.01$~eV,
$\Gamma^L_{1\sigma}=\Gamma^R_{2\sigma}=0.01$~eV,
$\Gamma^R_{1\sigma}=\Gamma^L_{2\sigma}=0$,
$U_1=U_2=U=1$~eV, $U_{12}=0.5$~eV, and $E_F=0.5$~eV.
As usually one sees pattern of blockaded and allowed transport regions.
However, here this pattern is more complicated than in the case of quantum dot.
Figure~\ref{fig_dqd_cut} demonstrates this pattern for a horizontal cut
of Fig.~\ref{fig_dqd} at source-drain voltage $V_{sd}/U=0.25$. 
Shown are current (a) and
probabilities to find the molecular subsystem in different occupation
states. 

\section{\label{conclude}Conclusion}
Approaches based on renormalization group technique (NRG, DMRG,
real time RG, etc.) are routinely used to treat quantum impurity
systems, where e.g. correlations between localized impurity
(quantum dot or molecule) and delocalized states (contacts)
leads to Kondo effect. Mostly these approaches were applied to
description of strongly correlated systems in equilibrium.
Description of transport is more problematic, since one needs
to know spectral function at finite temperature, where all excitations
may contribute.\cite{Bulla_review} Nevertheless first approaches
dealing with transport started to appear as
well.\cite{Bulla_review,Schoeller,Dagotto,Han,Cornaglia,Anders}
The main complication with implementation of these methods (besides DMRG)
for ab initio calculation is their complexity, so that all the calculations
done so far are restricted to simple models only.
In the case of molecular junctions most of ab initio calculations
done today are performed at the mean mean field level of treatment
with effective single particle orbitals used in place of molecular
states. Such approach clearly breaks down in the resonance tunneling
regime, where actual reduction/oxidation of the molecule leading
to corresponding electronic and vibrational structure change become
possible. Necessity of treating this regime in the language of
many-body molecular states, thus incorporating on-the-molecule correlations,
was realized and first approaches like e.g. generalized master
equation approach\cite{Datta_fseq,Datta_ph} where proposed.
Here we generalize this consideration by incorporating
many-body molecular states language into nonequilibrium Green's
function framework.
The main formal problem here is that many-body states language makes
the Wick's theorem inapplicable, and thus standard nonequilibrium diagrammatic
techniques can not be used. A workaround based on functional derivative
equation-of-motion technique for Hubbard operator GFs was developed 
by Sandalov and coworkers~\cite{Sandalov_ijqc} for equilibrium case.
The method so far has been applied to model
elastic transport through quantum dots\cite{FES_prl,F_prb72,SN_prb75} 
and lowest states of double quantum dots,\cite{FES_pn,FE_jpcm,F_prb69,FE_prb70} 
and is completely ignored in the molecular electronics community.
Here we employ the approach to deal with inelastic 
transport through molecular junctions in nonequilibrium atomic limit.
We formulate the method within basis of charged states of the molecule.
We demonstrate its ability to deal with transport situation in the
language of these states (rather than effective single-electron orbitals),   
as well as take into account charge-specific normal modes as well as 
nonadiabatic couplings. Capabilities of the technique are illustrated 
with simple model calculations of transport through quantum dot
and double quantum dot.  Extension to realistic calculations is the goal of
our future research.

\begin{acknowledgments}
M.G. is indebted to Igor Sandalov for numerous illuminating
discussions, and thanks Karsten Flensberg, 
Jonas Fransson and Ivar Martin for helpful conversations.
M.G. gratefully acknowledges the support of the UCSD Startup Fund and 
LANL Director's Postdoctoral Fellowship.
A.N. thanks the Israel Science Foundation, the US-Israel Binational
Science Foundation and the German-Israel Foundation for financial support.
M.R. thanks the NSF/MRSEC for support, through the NU-MRSEC.
This work was performed, in part, at the Center for Integrated
Nanotechnologies, a U.S. Department of Energy, Office of Basic Energy
Sciences user facility.  Los Alamos National Laboratory,
an affirmative action equal opportunity employer, is operated by
Los Alamos National Security, LLC, for the National Nuclear
Security Administration of the U.S. Department of Energy under
contract DE-AC52-06NA25396.
\end{acknowledgments}

\end{document}